\documentclass[a4paper,prb,amsmath,amssymb,twocolumn]{revtex4}
\usepackage{graphicx}
\usepackage{color}

\newcommand{\tr}{\text{Tr}\!}
\newcommand{\ex}[1]{\left\langle #1\right\rangle}
\newcommand{\kb}[2]{|#1\rangle\langle#2|}
\newcommand{\dg}[1]{#1^{\dagger}}
\newcommand{\ket}[1]{|#1\rangle}
\newcommand{\bra}[1]{\langle#1|}
\newcommand{\lket}[1]{|#1\rangle\!\rangle}
\newcommand{\lbra}[1]{\langle\!\langle#1|}
\newcommand{\lbk}[2]{\langle\!\langle#1|#2\rangle\!\rangle}
\newcommand{\liou}{\mathcal{L}}
\newcommand{\eVds}{eV_{\rm ds}}
\newcommand{\gext}{\gamma_{\rm ext}}

\newcommand{\mel}[3]{\lbra{l_0}#1\lket{r_{#3}}\!\lbra{l_{#3}}#2\lket{r_0}}

\newcommand{\glin}{\gamma_{\rm lin}}

\begin{document}

\title{Spectral properties of a resonator driven by a superconducting single-electron transistor}

\author{T. J. Harvey}
\author{D. A. Rodrigues}
\author{A. D. Armour}
\affiliation{School of Physics and Astronomy, University of
Nottingham, Nottingham, NG7 2RD, United Kingdom}

\begin{abstract}
We analyze the spectral properties of a resonator coupled to a
superconducting single electron transistor (SSET) close to the
Josephson quasiparticle resonance. Focussing on the regime where
the resonator is driven into a limit-cycle state by the SSET, we
investigate the behavior of the resonator linewidth and the energy
relaxation rate which control the widths of the main features in the
resonator spectra. We find that the linewidth becomes very narrow in
the limit-cycle regime, where it is dominated by a slow phase
diffusion process, as in a laser. The overall phase diffusion rate is determined by a
combination of direct phase diffusion and the effect of amplitude fluctuations which affect the
phase because the resonator frequency is amplitude dependent. For sufficiently strong couplings
we find that a regime emerges where the phase diffusion is no longer
minimized when the average resonator energy is maximized.
 Finally we show that the current
noise of the SSET provides a way of measuring both the linewidth and
energy relaxation rate.
\end{abstract}

\maketitle

\section{Introduction}

When a resonator is coupled to a superconducting single-electron
transistor (SSET) tuned close to the Josephson quasiparticle (JQP)
resonance the flow of charges can either damp the resonator motion,
potentially leading to cooling,\cite{clerk:05,blencowe:05} or pump
it\cite{bennett:06,rodrigues:07a,marthaler:08,harvey:08,andre:09,ashhab:09}
leading to laser-like states of self-sustaining oscillation,
depending on the choice of operating point. Recent experiments using
a superconducting stripline resonator coupled capacitively to the
SSET island demonstrated a laser effect~\cite{astafiev:07} whilst
the opposite effect of cooling was demonstrated using a
nanomechanical resonator.~\cite{naik:06} Similar cooling effects and
laser like instabilities occur close to other transport resonances
in the SSET-resonator system,\cite{clerk:05,bennett:06,koerting:09}
as well as in apparently rather different systems such as a driven
optical cavity coupled parametrically to a mechanical
resonator.\cite{kippenberg:08,favero:09,marquardt:09} Furthermore,
useful analogies can be made\cite{rodrigues:07,you:07,marthaler:08}
with quantum optical systems such as the micromaser where a cavity
resonator interacts with a sequence of two level
atoms.\cite{englert:02}

\begin{figure}
    \centering
    \includegraphics[width=115pt]{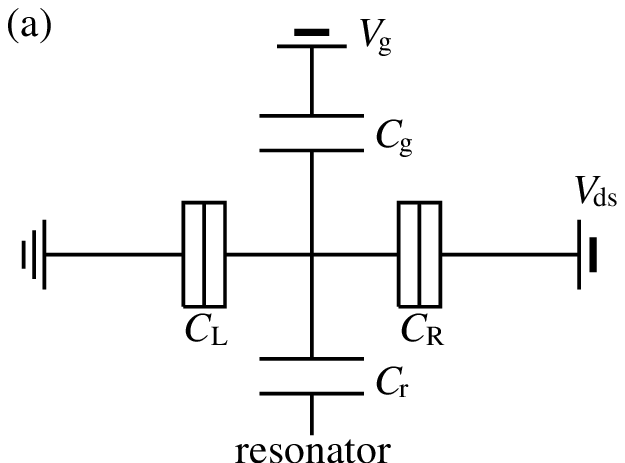}\hfill%
    \includegraphics[width=115pt]{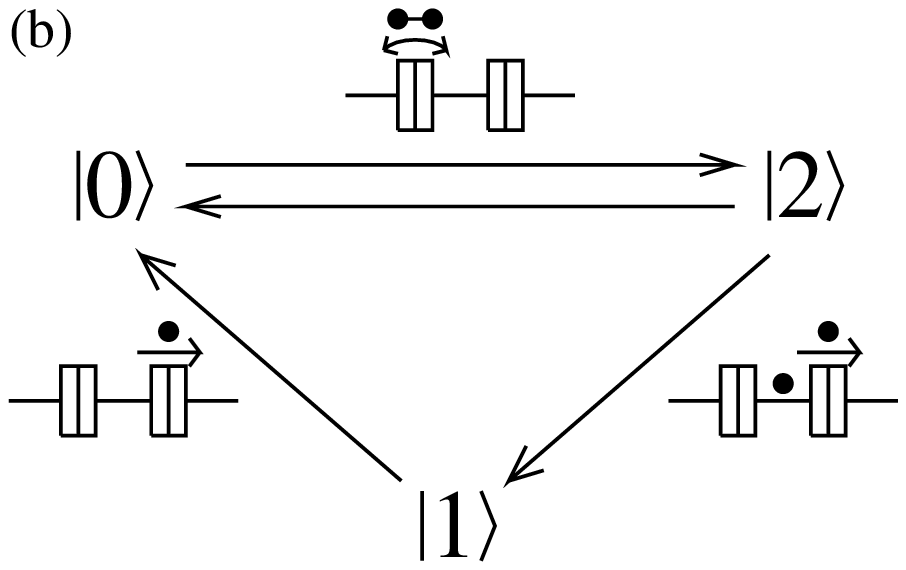}
    \caption{(a) Superconducting single electron transistor: two tunnel junctions and a gate capacitor $C_g$ form the SSET island to which the resonator is coupled capacitively. A drain source voltage, $V_{\rm ds}$, and gate voltage, $V_{\rm g}$, are used to tune the operating point of the device.  (b) JQP cycle: Cooper-pair tunneling at the left hand junction between island states $\ket{0}$ and $\ket{2}$ is interrupted by two quasiparticle tunneling events which take the island charge back to $\ket{0}$ via the $\ket{1}$ state.}
    \label{fig:set_jqp}
\end{figure}

The SSET is tuned to the JQP resonance by appropriate choices of the
drain source and gate voltages
applied~\cite{nazarov:09,aleshkin:90,choi:03} (see Fig.\
\ref{fig:set_jqp} for a schematic illustration of the JQP cycle).
Close to the resonance the charge dynamics of the SSET island is
similar to a driven two-level system coupled to a
bath.\cite{hauss:08,jaehne:08} Charge is transported through the
system via a combination of coherent tunneling of a Cooper-pair and
two successive quasiparticle decay processes. The center of the
resonance occurs when two states of the SSET island differing by a
single Cooper-pair, $|0\rangle$ and $|2\rangle$ have the same
charging energy. For operating points where the electrostatic energy
of state $|2\rangle$ is less than that of the $|0\rangle$ state the
SSET tends to emit energy to the
resonator.\cite{clerk:05,blencowe:05} The decay processes in the
SSET generate a current whose average value and fluctuations provide
a natural source of information about the dynamics of the
resonator.\cite{harvey:08,koerting:09}

For sufficiently strong coupling, the energy emitted to the
resonator can lead to a variety of different limit-cycle
states.\cite{clerk:05,bennett:06,rodrigues:07} The existence of the
limit-cycle states is shown clearly in the steady state properties
of the resonator's density matrix. However, information about the
important dynamical time-scales of the system such as the resonator
linewidth and energy relaxation rate is only obtained by going
beyond the steady-state of the system to examine the spectrum of
fluctuations that occur about this state.

In this paper we use a combination of numerical and analytical
methods to investigate the spectral properties of a resonator pumped
by a SSET tuned close to the JQP resonance. Direct numerical
evaluation of the relevant spectra allows us to obtain both the
energy relaxation rate and linewidth of the resonator. We find that
except within transition regions, the peaks in the resonator spectra
have a Lorentzian shape with widths that correspond to the real
parts of particular individual eigenvalues of the system. For weak
to moderate couplings the resonator linewidth behaves in a way that
is typical of self-sustained oscillators:\cite{lax:67} below
threshold it is simply half the energy relaxation rate of the resonator,
while above threshold, in the limit-cycle regime, it becomes much
narrower. We show that the linewidth in the limit-cycle regime is
set by the phase diffusion rate of the system in a way that is
similar to a laser.\cite{walls:94} For sufficiently strong coupling
we find that a more complex behavior emerges and the operating
points (of the SSET) at which the linewidth is narrowest no longer
correspond to the point where the resonator energy is maximized, an
effect which was also seen in a recent study of a very similar
system.\cite{andre:09}  We further show that the
full current noise spectrum of the SSET gives direct access to the
energy relaxation rate\cite{koerting:09} and, within a limit-cycle
state, the phase diffusion rate of the resonator.

This work is organized as follows. Sec.~\ref{sec:meth} is devoted to
the simple model we use to describe the SSET-resonator system and
the methods used in its solution. We begin by outlining the
Born-Markov master equation for the system, in which the evolution
of the reduced density matrix is given by a Liouvillian
superoperator, and then summarize the steady state properties of the
resonator. Next we describe the method used to calculate the
resonator spectra numerically. In Sec.~\ref{sec:res_linewidth} we
give the numerically calculated fluctuation spectra of the
resonator. Following a characterization of the features observed we
go on to show how the eigenvalues of the Liouvillian matrix can be
used to obtain the widths of the peaks observed in the spectra. The
calculation of the eigenvalues is significantly less challenging
numerically than a full calculation of the spectra, allowing us to
then investigate the behavior of the linewidth for a range of
coupling strengths. To understand the observed behavior of the
linewidth we make use of an analytical approximation in
Sec.~\ref{sec:anal_rate} and thereby link the eigenvalues to the
relevant physical processes in the system. Finally, in
Sec.~\ref{sec:cur_noise} we show that the current noise spectrum of
the system can be used to measure the linewidth and energy relaxation
rate of the resonator. The Appendix contains additional details of
the analytical calculations.

\section{Model\label{sec:meth}}

\subsection{Master Equation\label{sec:maseq}}

The Born-Markov master equation describing the evolution of the reduced density
matrix, $\rho(t)$, of the SSET and resonator in the vicinity of the
JQP resonance is given by,\cite{rodrigues:07,rodrigues:07a}
\begin{align}
    \dot{\rho}(t) &= -\frac{i}{\hbar}\left[H_{\rm co},\rho(t)\right] + \mathcal{L}_{\rm qp}\rho(t) + \mathcal{L}_{\rm d}\rho(t)\notag\\
    &= \mathcal{L}\rho(t).\label{eq:mas}
\end{align}
The first term describes the coherent evolution of the density
matrix under the Hamiltonian $H_{\rm co}$,
\begin{eqnarray}
    H_{\rm co} &=& \Delta E\kb{2}{2} - \frac{E_J}{2}\left(\kb{0}{2} + \kb{2}{0}\right) + \frac{p^2}{2m} \nonumber \\
    &&+ \frac{1}{2}m\Omega^2x^2 + m\Omega^2x_sx\left(\kb{1}{1} + 2\kb{2}{2}\right),\label{eq:hco}
\end{eqnarray}
where $\Delta E$ is the detuning from the JQP resonance and $E_J$ is
the Josephson coupling energy. The resonator has frequency $\Omega$,
effective mass $m$, momentum operator $p$, position operator $x$ and
$x_s$ parameterizes the SSET-resonator coupling strength.
 The second
and third terms of Eq.~\ref{eq:mas} describe the dissipative effects
of quasiparticle tunneling and the resonator's environment
respectively,
\begin{align}
    \mathcal{L}_{\rm qp}\rho(t) &= \Gamma\kb{0}{1}\rho(t)\kb{1}{0} + \Gamma\kb{1}{2}\rho(t)\kb{2}{1} \nonumber \\
    &- \frac{\Gamma}{2}\left\{\kb{1}{1} + \kb{2}{2},\rho(t)\right\},\label{eq:Lqp}\\
    \mathcal{L}_{\rm d}\rho(t) &= - \frac{\gext m\Omega}{\hbar}\left(n_{\rm ext}+\frac{1}{2}\right)\left[x,\left[x,\rho(t)\right]\right]\nonumber \\ &-\frac{i\gext}{2\hbar}\left[x,\left\{p,\rho(t)\right\}\right],
\end{align}
where $\Gamma$ is the quasiparticle tunneling rate and
$\{\cdot\,,\cdot\}$ is the anti-commutator. The effects of the
resonator's environment are parameterized by the external damping
rate $\gext$ and the thermal occupation number $n_{\rm ext}=({\rm
e}^{\hbar\Omega/k_{\rm B}T_{\rm ext}}-1)^{-1}$, with $T_{\rm ext}$
the temperature. We have neglected the (weak) dependence of $\Gamma$
on the position of the resonator and the difference in the tunneling
rates of the two quasiparticle decay processes.\cite{rodrigues:07}

For the numerical analysis of the system we use a Liouville space
representation,\cite{blum:96,briegel:93,jacob:04,flindt:04a,brandes:08}
following the notation introduced in Ref.~\onlinecite{harvey:08}.
The Liouvillian,\cite{djuric:06,harvey:08} $\liou$, appearing in
Eq.~\ref{eq:mas}, can be expressed in terms of an eigenvector
expansion,
\begin{equation}
    \liou = \sum_{p=0}^{\infty}\lambda_p\lket{l_p}\!\lbra{r_p},
    \label{eq:eigexpan}
\end{equation}
where $\lket{r_p}$ are the right hand eigenvectors, $\lbra{l_p}$ the
left hand eigenvectors and $\lambda_p$ the associated eigenvalues.
The steady-state density matrix in Hilbert space $\rho_{\rm ss}$ is
equivalent to the right hand eigenvector in Liouville space
corresponding to the eigenvalue $\lambda_0=0$, i.e. $\rho_{\rm
ss}\Leftrightarrow\lket{r_0}$. Multiplication on the left by
$\lbra{l_0}$ is the equivalent of the Hilbert space trace
operation,\cite{flindt:04a} $\lbk{l_0}{\rho(t)}={\rm Tr}[\rho(t)]$.

An appropriate truncation of the oscillator basis allows
Eq.~\eqref{eq:mas} to be solved numerically to find the relevant
eigenvalues and eigenvectors of $\liou$. We do this by using the
Matlab implementation of the ARPACK linear
solver.\cite{lehoucq:98} In order to use a large number of resonator
states we neglect the parts of the density matrix corresponding to
coherences involving the $\ket{1}$ state since these are decoupled
from the charge states of the system.\cite{rodrigues:07}
Additionally we make the approximation that the coherence between
resonator energy levels with a large separation in energy can be
neglected.\cite{harvey:08,koerting:09}

\subsection{Steady state behavior}

We now briefly review the steady-state properties of the resonator
as a function of the dimensionless coupling strength
$\kappa=\frac{m\Omega^2x_s^2}{eV_{\rm ds}}$ and the detuning $\Delta
E$ as this will provide important points of reference for the study
of the spectral properties in the following sections. For
concreteness we focus our analysis on the regime of a high frequency
resonator in comparison to the relaxation time of the SSET
($\Omega\gg\Gamma$), which is achievable in experiments using a
superconducting stripline resonator.\cite{astafiev:07} This
complements previous work on the low frequency
regime,\cite{clerk:05,bennett:06} although we note that our methods
are also suitable for a slow resonator. For this kind of resonator
the temperature can be sufficiently low for thermal effects to be
unimportant and so we take $n_{\rm ext}=0$ throughout. We also
choose to work in a regime where the Josephson coupling is
relatively weak compared to the quasiparticle decay rate and hence
we take $\Gamma=V_{\rm ds}/eR_J$. We use a value for the Josephson
energy of $E_J=1/16\,\eVds$, with a junction resistance
$R_J=h/e^2$. Working in this regime has the advantage that it is
also possible to develop an approximate analytical description of
the dynamics.\cite{rodrigues:07,rodrigues:08} Furthermore, when the
quasiparticle decay rate is relatively large this should be the
dominant source of dephasing for the SSET island charge and so we do
not need to consider additional environmental effects.

When $\Omega\gg\Gamma$ the interaction between the SSET and
resonator is quite weak except for near values of $\Delta E$ where
there is a matching of the eigenenergy of the SSET to the energy
level separation in the resonator,
\begin{equation}
    k\hbar\Omega = \pm\sqrt{\Delta E^2 + E_J^2},
    \label{eq:res_cond}
\end{equation}
where $k$ is a non-zero integer and the sign should match the sign
of $\Delta E$. For $\Delta E<0$ ($k<0$) these resonances correspond
to the transfer of energy from the SSET to the resonator. For
sufficient coupling the resonator is driven into states of
self-sustained oscillations.\cite{rodrigues:07a}

The steady state of the resonator is nicely described by the
distribution $P(n)={\rm Tr}[\ket{n}\bra{n}\rho_{\rm ss}]$, where
$\ket{n}$ is a Fock state of the resonator. The resonator is said to
be in a fixed point state when the $P(n)$ distribution has a single
peak at $n=0$. We define a well-defined limit-cycle state as
corresponding to a $P(n)$ distribution with a peak at $n\neq0$ and
additionally a small value for $P(0)$ (for the purposes of plotting
we choose `small' to mean $P(0)\leq1\times10^{-5}$). Finally we
define the transition region between the two states. In the regimes
studied here this occurs via a continuous transition, starting from
a fixed point state increasing the coupling $\kappa$ leads first to
a wider $P(n)$ distribution and then the peak in the distribution
moves away from $n=0$. We define the transition region as when the
peak in $P(n)$ is at $n\neq0$ but there is also substantial weight
at $P(0)$ (again we choose $P(0)>1\times10^{-5}$). Although this
particular definition of the transition region is of course somewhat
arbitrary, it nevertheless proves a useful indicator in what
follows.
 The resonator can display bistable (and multistable)
behavior in this system,\cite{rodrigues:07a} but here we focus only
on a parameter range where this is not the case and the $P(n)$
distribution only ever has a single peak.

\begin{figure}
    \centering
    \makebox[0pt][l]{\includegraphics{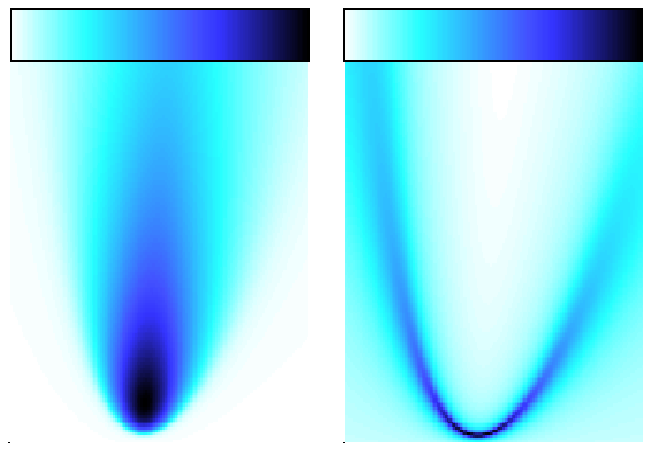}}%
    \includegraphics{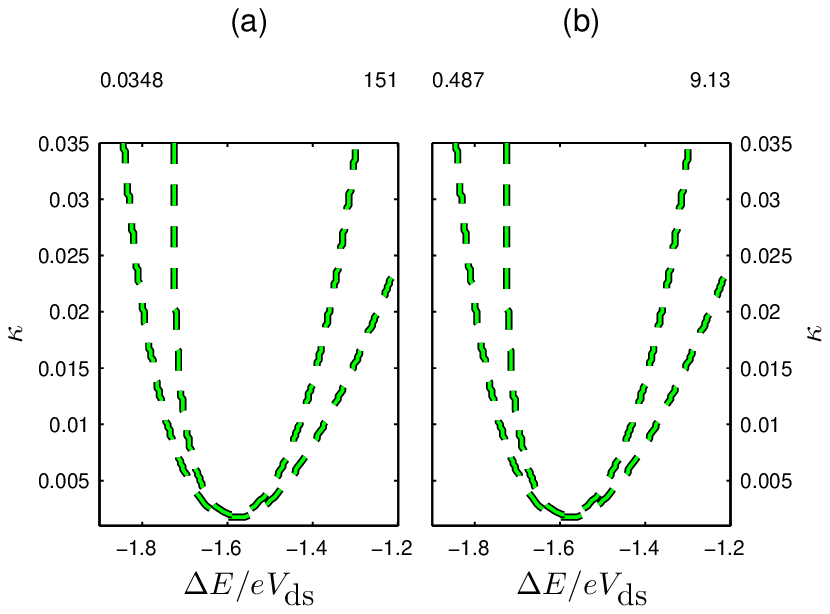}
    \caption{(Color online) (a) Average resonator energy $\ex{n}$ and (b) Fano factor $F_n$, for varying $\Delta E$ and $\kappa$,
    with $\Omega=10$ and $\gext=3\times10^{-4}$  (we adopt units such that $\Gamma=1$). Within the inner dashed line the resonator is in a limit-cycle state, between the two lines it is in the transition region (as defined in the text) and elsewhere it is in a fixed point state.}
    \label{fig:nav_Fn_de_kap}
\end{figure}

The steady state properties of the resonator,  $\ex{n}=\ex{\dg{a}a}$, where $a$ is the resonator lowering operator,
and $F_n=\ex{\bar{n}^2}/\ex{n}$, with $\bar{n}=n-\ex{n}$, are
shown in Figs.~\ref{fig:nav_Fn_de_kap}a and~\ref{fig:nav_Fn_de_kap}b
for varying detunings and coupling strengths. The plots are centered
around the $k=-1$ resonance [Eq.~\eqref{eq:res_cond}], which for our
parameters occurs at $\Delta E=-1.59\,\eVds$. Similar plots were
given in Ref.\ \onlinecite{harvey:08}, but here we go to higher
coupling. As the coupling is increased we see that $\ex{n}$
increases on resonance up to a maximum value for $\kappa\simeq0.005$
and then decreases. $F_n$ shows a large peak at the point where the
peak in the $P(n)$ distribution moves from $n=0$ to $n\neq0$ [see
Fig.~\ref{fig:nav_Fn_de_kap}b~\cite{harvey:08}]. The value of $F_n$
drops as the coupling is increased further and the system develops a
well-defined limit-cycle state. The value of $F_n$ drops below unity
(i.e.\ becomes sub-Poissonian) for sufficiently strong coupling
($\kappa \gtrsim 0.0075$ in Fig. \ref{fig:nav_Fn_de_kap}b).

%

\subsection{Noise spectra of the system\label{sec:noise_spec}}

The steady-state of a system only gives information about average
quantities. By also calculating noise spectra, further important
information about the dynamics of the system can be obtained. We
define the symmetrized noise spectrum for the two operators $b$ and
$\dg{b}$ as,
\begin{eqnarray}
    S_{b,\dg{b}}(\omega) & = & \lim_{t\to\infty}\int_{-\infty}^{\infty}d\tau\ex{\{\bar{b}(t+\tau),\bar{b}^{\dagger}(t)\}}e^{i\omega\tau}\label{eq:noise_spec_def}\\
  & =& 2\Re\lim_{t\to\infty}\int_0^{\infty}d\tau \ex{\{\bar{b}(t+\tau),\bar{\dg{b}}(t)\}}e^{i\omega\tau},
\end{eqnarray}
where $\bar{b}(t)=b(t)-\ex{b}$, $\ex{b}={\rm Tr}[b\rho_{\rm
ss}]$ and $\Re$ indicates the real part. The symmetrized noise spectrum has the property
$S_{b,b^\dag}(\omega)=S_{b^\dag,b}(-\omega)$. Note that we
consider fluctuations about the steady-state of the system
represented by the limit $t\to\infty$.

For system operators the correlation function can be evaluated by
using the quantum regression theorem (QRT). The QRT states that the
two-time correlation function can be written,\cite{walls:94}
\begin{equation}
    \lim_{t\to\infty}\ex{b(t+\tau)\dg{b}(t)} = \tr\left[be^{\mathcal{L}\tau}\dg{b}\rho_{\rm ss}\right],\qquad \tau\geq0.
    \label{eq:qrt_evol}
\end{equation}
In Liouville space this leads to,\cite{flindt:04a}
\begin{equation}
    S_{b,\dg{b}}(\omega) = 4\Re\lbra{l_0}\mathcal{B}\mathcal{R}(\omega)\dg{\mathcal{B}}\lket{r_0},
    \label{eq:nspec_sim}
\end{equation}
where $\mathcal{B}$ is the Liouville space symmetrized superoperator
whose relation to the Hilbert space operator $b$ is given by,
\begin{equation}
    \mathcal{B}\lket{\rho(t)} \Leftrightarrow \frac{1}{2}\left(b\rho(t) + \rho(t)b\right),
    \label{eq:sim_sup}
\end{equation}
and we define the pseudo-inverse of the Liouvillian,
\begin{equation}
    \mathcal{R}(\omega)=\mathcal{W}\left(-i\omega-\liou\right)^{-1}\mathcal{W},
    \label{eq:pinv}
\end{equation}
with $\mathcal{W}$ an operator that projects away from the
steady-state of the system, $\mathcal{W} \equiv 1 -
\lket{r_0}\!\lbra{l_0}$. The matrix formulation, given in
Eq.~\eqref{eq:nspec_sim}, is used to numerically evaluate the noise
spectra in the next section.

We can obtain helpful approximations and considerable insight into
how the resonator dynamics affect the noise by performing an
eigenfunction expansion of the Liouvillian [Eq.~\eqref{eq:eigexpan}]
in $\mathcal{R}(\omega)$ to obtain the alternative form for the
spectrum,
\begin{align}
    S_{b,\dg{b}}(\omega) &= 4\Re\left[\sum_{p=1}^{\infty}\frac{1}{-i\omega-\lambda_p}\lbra{l_0}\mathcal{B}\lket{r_p}\!\lbra{l_p}\dg{\mathcal{B}}\lket{r_0}\right]\notag\\
    &= -4\sum_{p=1}^{\infty}\frac{\Re[\lambda_p]\Re[m_B^p]-(\Im[\lambda_p]+\omega)\Im[m_B^p]}{\Re[\lambda_p]^2  +(\Im[\lambda_p]+\omega)^2},
    \label{eq:nspec_expan}
\end{align}
where $m_B^p=\mel{\mathcal{B}}{\dg{\mathcal{B}}}{p}$. In almost all the cases considered
here $m_B^p$ turns out to be real, and each term in the sum corresponds to a Lorentzian. However, in the case of the
current noise we find that $\Im[m_B^p]\neq 0$, leading to a somewhat more complex lineshape,\cite{rodrigues:08} a point which we discuss further
in Sec.~\ref{sec:cur_noise}.

A knowledge of the eigenvalues of the system tells us where we can
expect features in the spectra and how wide these features are. For
the two separate systems of a damped oscillator and the SSET near
the JQP resonance the eigenvalues are easily obtained. The
eigenvalues for an oscillator coupled to a thermal bath
are,\cite{englert:02}
\begin{equation}
    \lambda = -im\Omega - \left(l+\frac{1}{2}|m|\right)\gext \quad \begin{array}{l}l=0,1,2,\ldots\\m=0,\pm1,\pm2,\ldots\end{array}
\end{equation}
The eigenvalues of the SSET do not have such a simple analytic form.
For small $E_J$, the non-zero eigenvalues consist of a conjugate
pair with a real part $\sim -\Gamma/2$ and imaginary parts close to
the SSET frequency $\Omega_{\rm SSET}=\sqrt{\Delta
E^2+E_J^2}/\hbar$, with the others real and of order $-\Gamma$. In
the limit of zero coupling, the eigenvalues of the combined system
will just be the sum and difference of these eigenvalues.

\section{Spectral properties of the resonator\label{sec:res_linewidth}}

The first resonator spectrum we consider is
\begin{align}
    S_{a,\dg{a}}(\omega) &= \lim_{t\to\infty}\int_{-\infty}^{\infty}d\tau\ex{\{\bar{a}(t+\tau),\bar{\dg{a}}(t)\}}e^{i\omega\tau}\notag\\
    &= 4\Re\lbra{l_0}\mathcal{A}\mathcal{R}(\omega)\dg{\mathcal{A}}\lket{r_0},
\end{align}
where  $\mathcal{A}\lket{\rho(t)}\Leftrightarrow\frac{1}{2}\left(a\rho(t)+\rho(t)a\right)$.
The width of the peak that appears at the resonator frequency in
this spectrum is the resonator linewidth, $\gamma_{\Omega}$. For a
superconducting stripline resonator this spectrum can be inferred by
probing the field that leaks out of the resonator~\cite{walls:94}
(e.g.\ via capacitive coupling to a transmission
line~\cite{astafiev:07}), but here we will focus instead [in Sec.\
\ref{sec:cur_noise}] on how the current noise spectrum can be used
to obtain the resonator linewidth. For a resonator coupled only to a
thermal bath this spectrum consists of a single Lorentzian peak at
the frequency of the resonator, with a width given by half the
energy relaxation rate which in this case is
simply $\gamma_{\rm ext}/2$.

\begin{figure}
    \centering
    \includegraphics{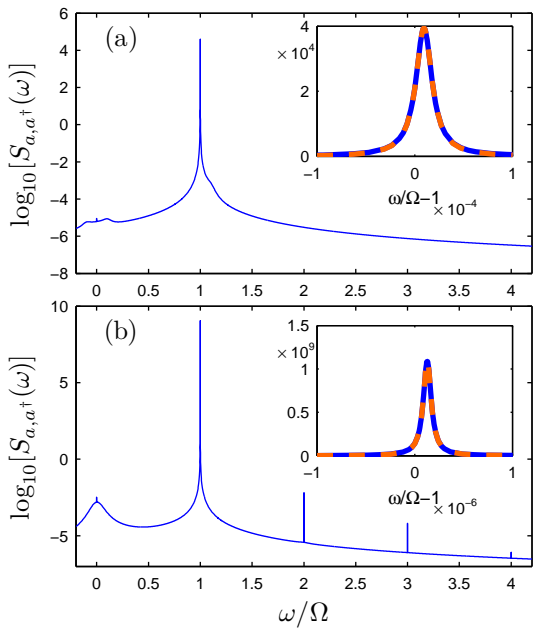}
    \caption{ (Color online) Resonator spectrum $S_{a,\dg{a}}(\omega)$ for $\kappa=0.003$ (other parameters are the same as in Fig.~\ref{fig:nav_Fn_de_kap}).
    (a) off-resonance $\Delta E=-1.75\,\eVds$, $\Omega_{\rm
SSET}=1.1\Omega$ on a $\log_{10}$ scale, (b) on-resonance $\Delta
E=-1.59\,\eVds$, $\Omega_{\rm SSET}=\Omega$ on a $\log_{10}$ scale.
The insets show the behavior around the peak at $\omega\simeq\Omega$
without logarithmic scaling (full lines) along with Lorentzian fits to the
peaks (dashed lines).}
    \label{fig:Saadg_spec}
\end{figure}

The behavior of $S_{a,\dg{a}}(\omega)$  for $\Delta E$ detuned away
from the center of the $k=-1$ resonance [Eq.~\eqref{eq:res_cond}]
(where the resonator is in a fixed point state) and at the resonance
(with $\kappa$ such that the system is just inside the region where
a well-defined limit-cycle exists) is shown in
Figs.~\ref{fig:Saadg_spec}a and~\ref{fig:Saadg_spec}b respectively.
For the off-resonant case, peaks are observed at $\omega=0$,
$\omega\simeq\Omega$, $\omega\simeq\Omega_{\rm SSET}$,
$\omega\simeq\Omega-\Omega_{\rm SSET}$ and
$\omega\simeq-\Omega+\Omega_{\rm SSET}$. When at resonance we also
observe the appearance of additional peaks at higher multiples of
the resonator frequency. Also note that the peaks involving the SSET
frequency have a much larger width since $\Gamma\gg\gext$.

The spectra are dominated by the peak near to the frequency of the
resonator, which is shown in detail in the insets. In each case the
position of the peak is shifted slightly from the bare frequency of
the resonator due to a renormalization arising from the interaction
with the SSET. In both cases the peak remains Lorentzian. By
comparing these two plots alone we see that the linewidth is much
narrower for the resonant case. The spectra are shown for a
relatively small value of $\kappa$, but for increased coupling
similar features are observed.

We also calculate the spectrum of energy fluctuations in the
resonator,
\begin{align}
    S_{n,n}(\omega) &= \lim_{t\to\infty}\int_{-\infty}^{\infty}d\tau\ex{\{\bar{n}(t+\tau),\bar{n}(t)\}}e^{i\omega\tau}\notag\\
    &= 4\Re\lbra{l_0}\mathcal{N}\mathcal{R}(\omega)\mathcal{N}\lket{r_0}.\label{eq:Snn_mat}
\end{align}
where
$\mathcal{N}\lket{\rho(t)}\Leftrightarrow\frac{1}{2}\left(n\rho(t)+\rho(t)n\right)$.
For a resonator alone this spectrum consists of a single peak at
$\omega=0$ with a width, $\gamma_n$, given by the energy relaxation
rate  ($\gext$ in this case) and height $4\ex{\bar{n}^2}/\gext$. For
the coupled system the peak at $\omega=0$ is found to be Lorentzian
to a very good approximation in both the limit-cycle and fixed point
regimes, but not within the transition region. We show in Sec.\
\ref{sec:anal_rate} that the width of this peak, $\gamma_n$, is
still given by the energy  relaxation rate of the resonator in both
the limit-cycle and fixed point states.

\begin{figure}
    \centering
    \includegraphics{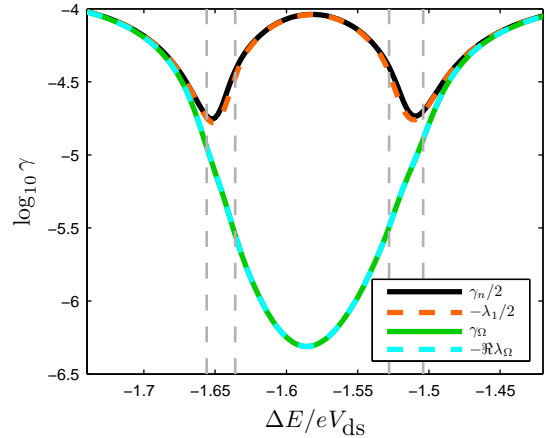}
    \caption{(Color online) Width of zero frequency peak in the $S_{n,n}(\omega)$ spectrum (upper curves) as determined from a
    Lorentzian fit ($\gamma_n/2$, solid line) and from the smallest non-zero eigenvalue ($-\lambda_1/2$, dashed line).
    The lower curves show the linewidth of the resonator as determined from a Lorentzian fit to $S_{a,\dg{a}}(\omega)$  ($\gamma_{\Omega}$, solid line) and from
    the eigenvalue ($-\Re\lambda_{\Omega}$, dashed line). $\kappa=0.003$ with other parameters the same as in Fig.~\ref{fig:nav_Fn_de_kap}.
    Vertical dashed lines indicate the transition region.}
    \label{fig:gn_gph_lorfit_eig}
\end{figure}

The behavior of the resonator linewidth, $\gamma_{\Omega}$, and the
rate $\gamma_n$ obtained from Lorentzian fits to the appropriate
spectral peaks are shown as a function of $\Delta E$ in Fig.\
\ref{fig:gn_gph_lorfit_eig}. The behavior far from resonance is
well-understood: the resonator remains in the fixed point state and
the SSET acts on it like an effective thermal
bath\cite{clerk:05,blencowe:05,rodrigues:08a,harvey:08} leading (for
$\Delta E<0$) to a reduction in the energy relaxation rate and an
increase in the effective frequency of the resonator. Thus one
expects that in this regime the resonator linewidth should be equal
to $\gamma_n/2$, as we find. Close to the center of the resonance
where the resonator is in a well-developed limit-cycle state there
is a strong suppression of the linewidth, whilst the energy
relaxation rate remains much larger. This behavior is precisely what
one expects for a self-sustained oscillator.\cite{lax:67}

The fact that a given spectrum can be expressed in terms of the
eigenfunction expansion of the Liouvillian as a sum of Lorentzians
[Eq.\ \ref{eq:nspec_expan}] suggests that where the spectral peaks
are found to be Lorentzian a single term in the expansion should
dominate and the peak width be given by a single eigenvalue.
Considering first the spectrum of energy fluctuations,
$S_{n,n}(\omega)$, we note that since this is the Fourier transform
of the correlation function $\ex{n(t+\tau)n(t)}$, we expect the
places where $\ex{\bar{n}^2}$ and $S_{n,n}(\omega)$ can be captured
by a single term in the eigenfunction expansion to be closely
related. In the fixed point and limit-cycle states (but not the
transition region) the first term in the eigenfunction expansion
describes the variance  in the resonator energy rather
well\cite{harvey:08}
$\ex{\bar{n}^2}\simeq\mel{\mathcal{N}}{\mathcal{N}}{1}$, where the
corresponding eigenvalue $\lambda_1$ is the smallest non-zero one.
The single term approximation to the spectrum is,
\begin{align}
    S_{n,n}(\omega) &\simeq 4\Re\left[\frac{1}{-i\omega-\lambda_1}\mel{\mathcal{N}}{\mathcal{N}}{1}\right]\notag\\
    &\simeq \frac{4(-\lambda_1)\ex{\bar{n}^2}}{\omega^2+\lambda_1^2}.
\end{align}
The validity of this expression is confirmed in Fig.\
\ref{fig:gn_gph_lorfit_eig} where we compare the eigenvalue
$\lambda_1$ and $\gamma_n$ and find excellent agreement except
within the transition region, where the $\Omega=0$ peak is not
well-described by a Lorentzian and further terms in the
eigenfunction expansion are required.

 For $S_{a,\dg{a}}$ it is expected
that the eigenvalue closest to $-i\Omega$ is the most important
(this is certainly true for the decoupled system). We denote this
eigenvalue $\lambda_{\Omega}$. For the eigenfunction expansion of
the steady-state quantity $\ex{\{\bar{a},\bar{\dg{a}}\}}$ we find
that (for fixed point and limit-cycle states and also in the
transition region for weak coupling) the term corresponding to the
$\lambda_{\Omega}$ eigenvalue dominates, i.e.\
$\frac{1}{2}+\ex{n}\simeq\mel{\mathcal{A}}{\dg{\mathcal{A}}}{\Omega}$.
Thus we expect that the spectrum around $\omega=\Omega$ should be
well approximated by,
\begin{equation}
    S_{a,\dg{a}}(\omega) \simeq \frac{4(-\Re\lambda_{\Omega})\left(\ex{n}+\frac{1}{2}\right)}{(\omega-\Omega_R)^2+[\Re\lambda_{\Omega}]^2},
    \label{eq:S_emm_expan}
\end{equation}
where $\Omega_R$ is the renormalized frequency of the resonator,
which is given by the imaginary part of the eigenvalue,
$\Omega_R=-\Im[\lambda_{\Omega}]$. The comparison of
$\gamma_{\Omega}$ with $\Re\lambda_{\Omega}$ in Fig.\
\ref{fig:gn_gph_lorfit_eig} shows convincing agreement.

\begin{figure}
    \centering
    \makebox[0pt][l]{\includegraphics{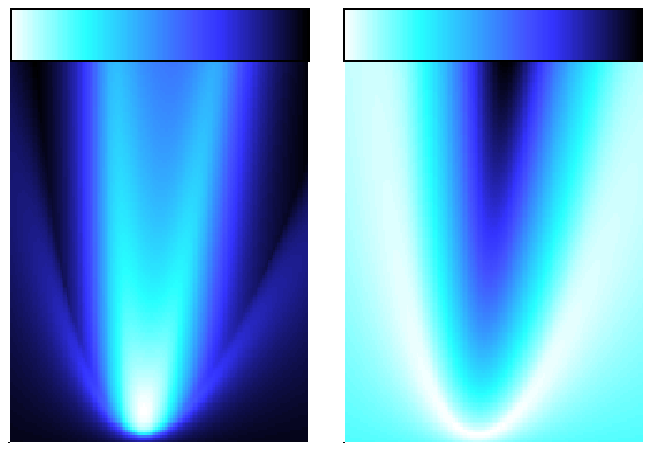}}%
    \includegraphics{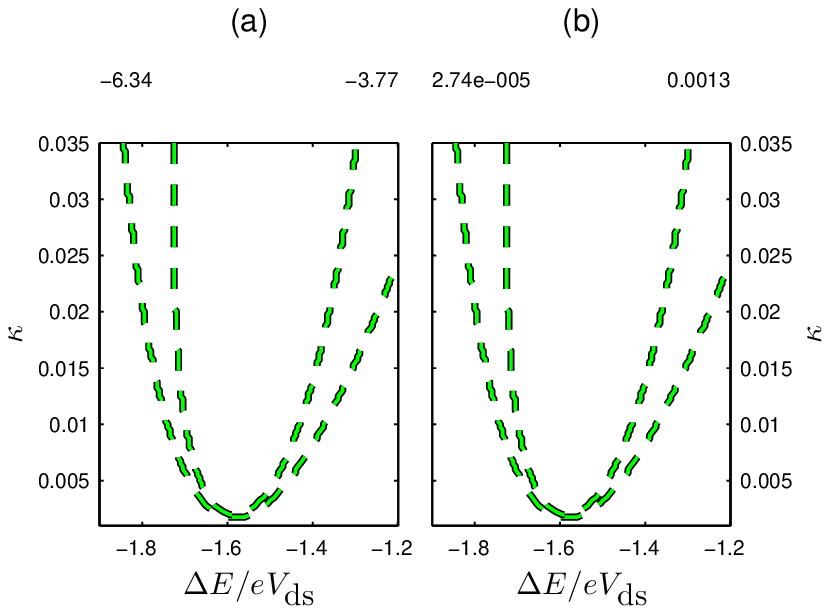}
    \caption{(Color online) Widths of spectral peaks calculated from the eigenvalues as a function of $\kappa$ and
     $\Delta E$. (a) $\log_{10}\left[\gamma_{\Omega}\right]$, the logarithm of the linewidth of $S_{a,\dg{a}}(\omega)$
    (b)$\gamma_n$, the width of the zero-frequency peak in $S_{n,n}(\omega)$,
    (other parameters and lines are as in Fig.~\ref{fig:nav_Fn_de_kap}).}
    \label{fig:gom_gn_de_kap}
\end{figure}

The calculation of the eigenvalues is much less numerically
intensive than the calculation of the full spectrum, allowing us to
calculate the linewidth for  a large range of parameters. In
Fig.~\ref{fig:gom_gn_de_kap}a we show the value of
$\gamma_{\Omega}$, as determined from the $\lambda_{\Omega}$
eigenvalue, for the same range of parameters as
Fig.~\ref{fig:nav_Fn_de_kap}. For weak coupling
$\gamma_{\Omega}$ decreases with increasing
$\ex{n}$ and hence the maximum in $\ex{n}$ corresponds to a minimum
in the linewidth. This behavior is analogous to that seen in a laser
where the linewidth above threshold is given by the rate of phase
diffusion,\cite{walls:94} $\gamma_{\phi}^{\rm laser} =
\frac{G}{8\ex{n}}$, where $G$ is the gain. However, a more complex behavior is apparent
close to resonance for $\kappa> 0.005$ where $\ex{n}$ decreases with increasing $\kappa$ (see Fig.\ \ref{fig:nav_Fn_de_kap}) and the
(single) maximum in $\ex{n}$ now corresponds to a local maximum in
$\gamma_{\Omega}$ with minima on either side.

The behavior of $\gamma_{\Omega}$ manifests itself in the height of
 peak at $\omega\sim \Omega$ in $S_{a,\dg{a}}(\omega)$, which from
Eq.~\eqref{eq:S_emm_expan}, is  $\simeq4\ex{n}/\gamma_{\Omega}$
($\ex{n}\gg1$ in the limit-cycle). The height of this peak as a function of $\Delta E$
initially shows a single maximum at $\Delta E=-1.59\,\eVds$, which
splits into a pair of maximas (with a minimum between) as the coupling is increased. Similar behavior
was observed in the emission spectrum of a stripline
resonator coupled to a SSET in a recent
experiment.\cite{astafiev:07}  However, this experiment was in a
rather
 different parameter regime to the one considered here ($E_J\gg \hbar\Gamma$) and in the
experiment there were accompanying features in the current, which we do
not obtain here. A double peak in the spectral maximum as a function of the detuning from resonance was also predicted in the emission
spectrum for a closely related system in Ref.~\onlinecite{andre:09}. In contrast to the linewidth, the energy relaxation rate $\gamma_n$ shown in
Fig.~\ref{fig:gom_gn_de_kap}b shows a similar behavior  to that shown in
Fig.~\ref{fig:gn_gph_lorfit_eig} for all the
values of $\kappa$ plotted.

\section{Origin of the peak widths\label{sec:anal_rate}}

We have shown that the widths of the peaks in the noise spectra are
related to the eigenvalues of the Liouvillian, and that when the
peaks are Lorentzian, the widths are given by a single eigenvalue
corresponding to the dominant noise process. In this section we
identify the eigenvalues with physical processes by comparing them
to analytic expressions that describe those processes.

We first consider $\gamma_n$, the width of the $\omega=0$ peak in
$S_{n,n}(\omega)$. We do not expect the correlation function of $n$
to depend on phase, and so in Section \ref{sec:res_linewidth} we
identified this term as the energy relaxation rate. This is
supported by the fact that this term reduces to $\gamma_{\rm ext}$ in
the zero-coupling limit. It is well known that when the resonator is
in a fixed point state, the energy relaxation rate is given by the
sum of the external damping and an additional effective damping due
to the SSET,\cite{blencowe:05, clerk:05} and indeed we find that
when the resonator is in a fixed point state, $\gamma_n$ is well
approximated by this total damping rate. We will now go on to show
that $\gamma_n$ also corresponds to the energy relaxation rate in
the limit-cycle.

In a limit-cycle, the effective damping can be generalized to obtain
an amplitude dependent
damping\cite{clerk:05,bennett:06,rodrigues:07,rodrigues:08a}
$\gamma_{\rm eff}(E)$ where $E\approx
\ex{n}$  is the average energy. The calculation, which is described in Ref.\ \onlinecite{rodrigues:08a}, proceeds by deriving a set of Langevin equations for the SSET and resonator operators.
Assuming then that the resonator energy relaxes much more slowly than the SSET charge, the problem can be separated into two parts (though at the cost of neglecting some of the SSET-resonator correlations). Firstly the SSET charge dynamics can be solved treating the resonator as an harmonic drive with a fixed amplitude. The driven charge response as a function of the resonator amplitude is then used to obtain an effective Langevin equation for the resonator alone.\cite{rodrigues:08a} Writing down the corresponding equation for the average amplitude of the resonator leads immediately to the effective damping, an expression for which (valid in the limit $E_J\ll\Gamma$) is given in the
appendix.
%
The average energy of the resonator obeys the equation of motion,
\begin{equation} \frac{dE}{dt} =
-\gamma_T(E)E,
\end{equation}
where $\gamma_T(E)=\gamma_{\rm ext}+\gamma_{\rm eff}(E)$ is the
total damping rate. Although this equation is nonlinear, for small
fluctuations about a stable limit-cycle with average energy $E_0$
(given by $\gamma_T(E_0)=0$) we can linearize this equation to obtain,
\begin{align}
    \frac{dE}{dt}     &\simeq -\glin\left(E-E_0\right)\notag\\
    \glin &= E_0\left.\frac{d\gamma(E)_T}{dE}\right|_{E=E_0},
\end{align}
where $\glin$ is the linearized damping, which gives the energy
relaxation rate of the resonator near to a stable limit-cycle
solution. The same approach also allows us to calculate the
renormalized frequency of the resonator, $\Omega_R$ (as we discuss
in the appendix).

\begin{figure}
    \centering
    \includegraphics{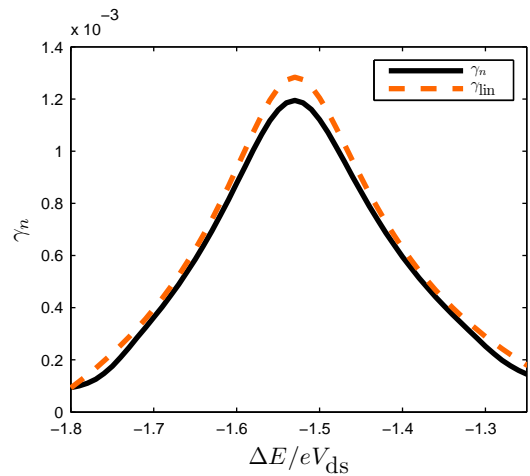}
    \caption{(Color online) Comparison of the width of the zero frequency peak in
    $S_{n,n}(\omega)$, as given by the eigenvalue $\gamma_n(\Delta \tilde{E})$ (solid line), with the energy relaxation
    rate $\glin(\Delta E)$ (dashed line).
    Here $\kappa=0.03$, with other parameters the same as in Fig.\ \ref{fig:nav_Fn_de_kap}.}
    \label{fig:gn_glin}
\end{figure}

The calculation of $\gamma_{\rm eff}(E)$
requires\cite{rodrigues:07,rodrigues:08a} approximations
in which certain correlations between the resonator and the SSET are dropped. Not surprisingly, this
approach does not capture all of the coupled dynamics of the system and
in particular it does not describe the small shift in the SSET
frequency which arises due to the coupling to the resonator.  This
SSET frequency shift is apparent in the plot of $\ex{n}$ in
Fig.~\ref{fig:nav_Fn_de_kap}a, where there is a clear change in the
position of the peak as the coupling is increased (since this shift is much larger than
than $\Omega_R-\Omega$, it can be attributed to a shift in $\Omega_{SSET}$). The shifted
frequency of the SSET can be obtained from a calculation which
includes more of the SSET-resonator correlations (for
example from the shift in the SSET eigenvalues within the second
order mean field equations described in Ref.\
\onlinecite{harvey:08}), or simply from the precise location of the
peaks in the current noise spectrum of the SSET (Section
\ref{sec:cur_noise}). In Fig.~\ref{fig:gn_glin} we compare
$\gamma_n$ obtained numerically (from the Liouvillian eigenvalue
$\lambda_1$)  to $\gamma_{\rm lin}$ plotted against $\Delta E$ renormalized
to take into account the shift in $\Omega_{SSET}$ (for low-$E_J$ and high-$\Omega$, the required renormalization is well
approximated by $\Delta \widetilde E\approx \Delta E+2\kappa
eV_{ds}$). We find good agreement for these parameters which
confirms that our identification of $\gamma_n$ as the energy
relaxation rate is indeed valid. The small difference between the
curves in Fig.~\ref{fig:gn_glin} arises because the expression for
$\gamma_{\rm lin}$ is strictly only valid in the small $E_J$ limit; we
find that a smaller value of $E_J$ improves the agreement.


We now consider $\gamma_\Omega$, the width of the $\omega=\Omega$
peak in $S_{a,\dg{a}}(\omega)$. In the fixed point regime, we find
that $\gamma_\Omega\approx \gamma_n/2$, i.e.\ this width is also
determined by the energy relaxation rate --- indicating that the
dominant fluctuations of $a$ are energy fluctuations in this regime. However, in
the limit-cycle regime where the energy fluctuations are small there
is an additional process by which $a$ can fluctuate, namely
phase diffusion. The linewidth in a laser is determined by this
phase diffusion rate,\cite{lax:66,walls:94} and we shall now show
that this is also the case for the SSET-resonator system.

Mapping the effective Langevin equation for the resonator\cite{rodrigues:08a}
onto a c-number equation,\cite{scully:97} we then follow the approach of
Lax,\cite{lax:66,lax:67} and derive effective equations of motion for the
phase and amplitude. The calculation follows very closely that described in Ref.\ \onlinecite{rodrigues:09} where
the phase diffusion rate of a resonator coupled to a driven cavity
is obtained. The effective amplitude and phase equations contain
terms representing fluctuations arising from both the environment and the
the SSET island charge. These fluctuations (calculated in
Ref.~\onlinecite{rodrigues:08a} and summarized in the appendix) lead
to effective diffusion terms for the amplitude ($D_{\rm eff}^-$) and
phase ($D_{\rm eff}^+$) of the resonator about the limit-cycle,
where the effective diffusion constants are valid on timescales long
compared to the resonator period, but short compared to the damping
rate.

We find\cite{rodrigues:09} that as well as the direct phase
diffusion $D^+_{\rm eff}$, we must also take into account another
mechanism of phase diffusion. The frequency shift
$\Omega'=\Omega_R-\Omega$ depends on the amplitude, so any change in
amplitude will lead to a change in frequency.\cite{bennett:06} This
means that a term describing amplitude fluctuations appears in
the equation of motion of phase, and thus there are both direct and
indirect contributions to the phase diffusion.

\begin{figure}
    \centering
    \includegraphics{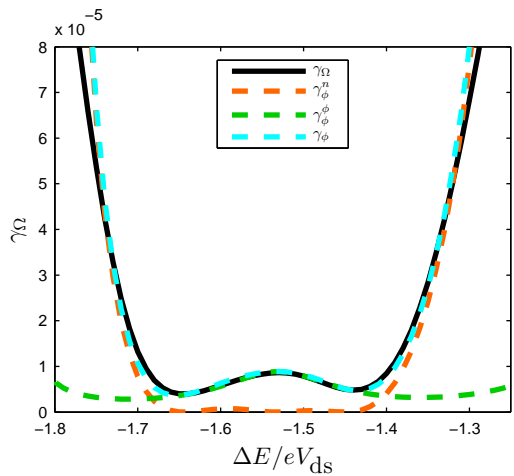}
    \caption{(Color online) Comparison of the linewidth $\gamma_{\Omega}$ of $S_{a,\dg{a}}(\omega)$ obtained from the $\lambda_{\Omega}$ eigenvalue (plotted as a function of $\Delta \tilde{E}$) with the analytic expression $\gamma_{\phi}=\gamma_{\phi}^{\phi}+\gamma_{\phi}^n$ (plotted as a function of $\Delta {E}$). Also shown are the
    direct ($\gamma_{\phi}^{\phi}$) and indirect ($\gamma_{\phi}^n$) phase diffusion contributions. Here $\kappa=0.03$, with other parameters the same as in Fig.\ \ref{fig:nav_Fn_de_kap}.}
    \label{fig:gom_gomlin}
\end{figure}

We combine the  terms arising from direct phase diffusion and the contribution from amplitude fluctuations (neglecting cross-correlations) to obtain the total linewidth
$\gamma_\phi=\gamma_\phi^\phi+\gamma_\phi^n$,
\begin{equation}
    \gamma_{\phi} = \frac{D_{\rm ext}+D_{\rm eff}^+(E)}{4\ex{n}} + \left(\frac{\Omega_{\rm lin}}{\glin}\right)^2 \frac{ D_{\rm ext}+D_{\rm eff}^-(E)}{\ex{n}}
    \label{eq:gphi}
\end{equation}
where $\Omega_{\rm
lin}=E_0\left.\frac{d\Omega'(E)}{dE}\right|_{E=E_0}$ is the
frequency shift linearized about the limit-cycle, and
$D_{\rm ext}=\frac{\gamma_{\rm ext}}{2}$ at zero temperature. In
Fig.~\ref{fig:gom_gomlin} we compare $\gamma_\phi$ with the
numerically obtained result for the linewidth from the eigenvalue,
$\gamma_\Omega$. We see that near to the center of the resonance the
indirect diffusion term tends to zero, but it contributes
significantly when the system is not exactly on resonance.

Looking at Fig.~\ref{fig:gom_gomlin} we see that the double-peak structure seen in $\gamma_{\phi}$ at strong couplings arises from the combination of the maximum in $\gamma_{\phi}^{\phi}$ at the center of the resonance and the strong increase in $\gamma_{\phi}^{n}$ off-resonance. At weaker couplings the limit-cycle regime becomes much narrower as a function of $\Delta E$. Here we find that the combination of a less pronounced maximum in $\gamma_{\phi}^{\phi}$  and a much sharper minimum in $\gamma_{\phi}^{n}$ together lead to a single minimum in $\gamma_{\phi}$.

\section{Current noise\label{sec:cur_noise}}

In this section we investigate the extent to which features in the
resonator noise spectra from Sec.~\ref{sec:res_linewidth} manifest
themselves in the current noise. The low frequency limit of the
current noise in this system was discussed in
Ref.~\onlinecite{harvey:08}, but here our interest is in the
dynamical timescales which become imprinted in the widths of the
peaks in the finite frequency current noise.  The finite frequency
current noise through the SSET can be split into contributions from
the current noise at the two junctions and the charge noise of the
island,\cite{mozyrsky:02}
\begin{equation}
    S_{I,I}(\omega) = \frac{1}{2}S_{I_L,I_L}(\omega) + \frac{1}{2}S_{I_R,I_R}(\omega) - \frac{1}{4}\omega^2S_{Q,Q}(\omega),
\end{equation}
where we have assumed equal junction capacitances. We choose the
Cooper pair tunneling to take place at the left hand junction. To
calculate the charge noise and the current noise at the left hand
junction we can again use the quantum regression theorem as
described in Sec.~\ref{sec:noise_spec},
\begin{align}
    S_{I_L,I_L}(\omega) &= 4\Re\lbra{l_0}\mathcal{I}_L\mathcal{R}(\omega)\mathcal{I}_L\lket{r_0},\notag\\
    S_{Q,Q}(\omega) &= 4\Re\lbra{l_0}\mathcal{Q}\mathcal{R}(\omega)\mathcal{Q}\lket{r_0},
\end{align}
where the current and charge operators are defined as,
\begin{align}
    I_L &= ie\frac{E_J}{\hbar}\Big(\kb{2}{0}-\kb{0}{2}\Big),\notag\\
    Q &= e\Big(\kb{1}{1} + 2\kb{2}{2}\Big).
\end{align}
and the superoperator forms are defined in the same manner as
Eq.~\eqref{eq:sim_sup}. For the current noise at the right hand
junction a counting variable approach can be used.\cite{flindt:05}
The resulting expression is very much the same but with the addition
of a self-correlation term,
\begin{equation}
    S_{I_R,I_R}(\omega) = 2e\lbra{l_0}\mathcal{I}_R\lket{r_0} + 4\Re\lbra{l_0}\mathcal{I}_R\mathcal{R}(\omega)\mathcal{I}_R\lket{r_0},
\end{equation}
where the current super operator for the right hand junction is
defined,
\begin{equation}
    \mathcal{I}_R\lket{\rho(t)} = e\Gamma\Big(\kb{0}{1}\rho(t)\kb{1}{0} + \kb{1}{2}\rho(t)\kb{2}{1}\Big).
\end{equation}
We give results in terms of the current Fano factor,
$F_I(\omega)=S_{I,I}(\omega)/2e\ex{I}$. For the SSET alone the
current noise spectrum, for these parameters, consists of a large
peak at $\omega\simeq\Omega_{\rm SSET}$ and a much smaller peak at
$\omega=0$.

\begin{figure}
    \centering
    \includegraphics{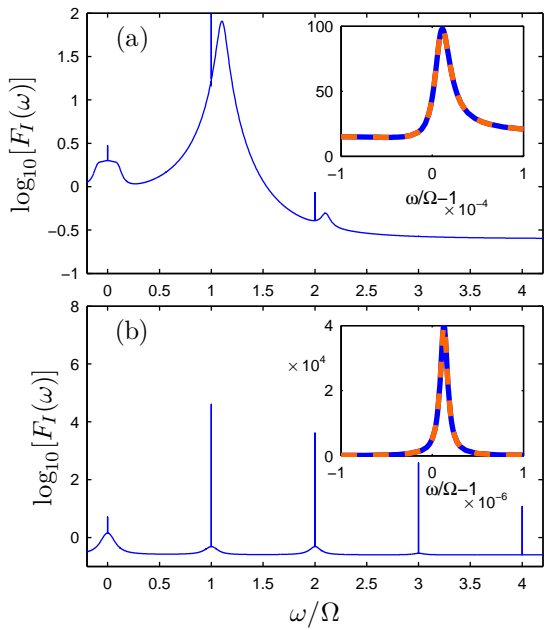}
    \caption{(Color online) Finite frequency current noise spectrum for $\kappa=0.003$
    with the other parameters the same as in Fig.~\ref{fig:nav_Fn_de_kap}.
    (a) Off-resonance $\Delta E=-1.75\,\eVds$, (b) on-resonance for $\Delta E=-1.59\,\eVds$.
    The insets show expansions of the narrow peaks at $\omega\simeq\Omega$ without logarithmic scaling (full lines) along with fits
    to the peaks (dashed lines) with the form of Eq.~\ref{eq:nspec_expan}, inset (a), and Lorentzian form, inset (b).}
    \label{fig:FI_spec}
\end{figure}

In Figs.~\ref{fig:FI_spec}a and~\ref{fig:FI_spec}b we show the
current noise spectra for the same off-resonant and resonant
parameters used in Figs.~\ref{fig:Saadg_spec}a
and~\ref{fig:Saadg_spec}b. The SSET current has a non-linear dependence on the resonator position\cite{harvey:08} leading to peaks at higher harmonics\cite{flindt:05a} of $\Omega$. In the off-resonant regime the resonator state consists of Gaussian fluctuations about a fixed point and only the $\omega\simeq2\Omega$ harmonic is seen in the current noise\cite{armour:04,doiron:06} and it is very much weaker than the peak at $\omega\simeq\Omega$. On-resonance, the resonator motion consists of oscillations of a (relatively) much larger amplitude and hence leads to clearly visible peaks at higher harmonics of $\Omega$.

It is interesting to note that for the peak near $\omega\simeq\Omega$ the off-resonant current noise
spectrum displays an important difference from the
resonator spectra: as is shown in the inset
(Fig.~\ref{fig:FI_spec}a) the peak is not of a Lorentzian
form. The non-Lorentzian shape of this peak in the fixed point regime can be
understood\cite{rodrigues:08} as arising from an extra element of correlation between the SSET and resonator:
the SSET current picks up the fluctuations in
the resonator motion which were originally driven by the SSET charge
motion (rather than a totally uncorrelated external bath).\cite{rodrigues:08} This effect is rapidly reduced with
increasing external damping and temperature. Note that in the limit-cycle (resonant) region the peak at $\omega\simeq\Omega$ is again a Lorentzian.

Nevertheless, we find that the non-Lorentzian peak at $\omega\simeq\Omega$ in the
fixed point regime can still be fitted by a term in the eigenfunction expansion
Eq.~\ref{eq:nspec_expan} (but with a complex matrix element $m_B^p$), and
hence
the
parameter describing the width, $\Re\lambda_{\Omega,I}$, can be
obtained over the whole range of $\Delta E$. Comparing
$\Re\lambda_{\Omega,I}$ obtained in this way with the linewidth
$\gamma_{\phi}$, and comparing the width of the $\omega=0$ peak in $S_{I,I}(\omega)$
with $\gamma_n$ for the parameters in Fig.\
\ref{fig:gn_gph_lorfit_eig} we find in both cases that the pairs of
curves overlay one another (the resulting plot thus effectively duplicates that shown in Fig.\
\ref{fig:gn_gph_lorfit_eig} and hence is not shown here).  Hence the current noise
provides a direct measure of the spectral properties of the
resonator spectra and, in particular, gives access to the energy
relaxation and phase diffusion rates.

\section{Conclusions}

We have studied the noise properties of a resonator coupled to a
SSET in the vicinity of the JQP resonance, focussing our analysis on
the case of a resonator whose frequency is large compared to the
quasiparticle decay rate. We analyze the resonator spectrum that
includes fluctuations in both the resonator energy and phase,
$S_{a,\dg{a}}(\omega)$, as well as the spectrum of energy
fluctuations $S_{n,\dg{n}}(\omega)$. The main feature in the
spectrum $S_{a,\dg{a}}(\omega)$ is a large peak at the resonator
frequency, $\omega\approx\Omega$ while $S_{n,\dg{n}}(\omega)$ is
dominated by a peak at $\omega=0$. We find that outside transition
regions, these peaks are Lorentzian with widths, $\gamma_{\Omega}$
(the linewidth) and $\gamma_n$, respectively, which are each
controlled by a single eigenvalue of the Liouvillian. Examining the
behavior of these eigenvalues for a range of coupling strengths, we
find that for large enough coupling $\lambda_{\Omega}$ has a double
minimum as a function of the detuning from resonance. An analysis of
the coupled dynamics of the system allows us to identify the peak
width $\gamma_n$ with the energy relaxation rate and $\gamma_\Omega$
(within the limit-cycle regime) with the phase diffusion rate. We
show that the double minimum in the linewidth arises from a
combination of direct and indirect phase diffusion. Finally, we show
that these rates can be extracted from the current noise, and thus
the current noise provides a direct measure of the key dynamical time
scales of the resonator.

\section*{Acknowledgements}

This work was supported by EPSRC under Grants EP/D066417/1 and
EP/C540182/1.

\appendix

\section{Energy dependent damping and diffusion}

In this appendix we present further details on the analytical approximations used to obtain the
effective damping and diffusion rates. The
details of the methods used in the calculations (and the approximations required)
can be found in Refs.\
\onlinecite{rodrigues:08a} and \onlinecite{rodrigues:09}. Although the expressions obtained are only valid for
small Josephson energy, $E_J \ll \hbar\Gamma$, they are
valid for all values of $\Omega/\Gamma$, i.e.\ for both fast and slow
resonators. The basis of the calculation is that the resonator energy is
assumed to relax very slowly compared to both the resonator period and the quasiparticle
tunneling rate, so that its effect on the SSET charge can be approximated as a harmonic drive. The
behavior of the SSET under the influence of this drive is then
found, and the result then fed back to obtain a coarse-grained
equation of motion for the resonator.

The effective damping, $\gamma_{\rm eff}$, and frequency shift, $
\Omega'=\Omega_R-\Omega$, for a resonator with average (real) amplitude $\sqrt{E}$ are given by,\cite{rodrigues:08a}
\begin{align}
\frac{\gamma_{\rm eff}(E)}{2}+i\Omega'(E)&=-i\frac{x_s \Omega \pi^2
E_J^2 \Gamma^2}{x_q (eV_{ds})^2
2\sqrt{E}}\frac{(3\Gamma+2i\Omega)}{(\Gamma+i\Omega)^2}\nonumber \\
&\times \sum\limits_n
J_{n}(z)J_{1-n}(z)\left(\frac{1}{h_{n}}-\frac{1}{h_{-n}^*}\right)
\end{align}
where $x_q=\sqrt{\hbar/(2m\Omega)}$, the Bessel function of the first kind
$J_n(z)$ is a function of the scaled amplitude $z=2\sqrt{
E}x_s/x_q$ and we define $h_n=\frac{\Gamma}{2} +i (\Omega n + \Delta
E/\hbar)$.

In a similar way, the fluctuations of the charge on the
SSET\cite{rodrigues:08a} lead to effective diffusion
constants for the amplitude and phase of the resonator.\cite{rodrigues:09} The details of the calculation are
rather involved, and the calculation follows the same approach as that given in Ref.\ \onlinecite{rodrigues:09} hence here we merely summarize the results. The total
diffusion rate is given by $D_{\rm
eff}^\pm=\frac{E_J^2\Gamma^4}{(eV_{ds})^2}\pi ^3\Omega\kappa
\Re\left(D_1^\pm+ D_2^\pm+ D_3^\pm\right)$, where $D_{\rm
eff}^+$ refers to the phase diffusion, and the $D_{\rm
eff}^-$
terms to the amplitude diffusion. The first term is,
\begin{align}
D_1^\pm &= \frac{1}{2}\sum\limits_n \frac{\left|J_{n+1}(z)C_1\pm
J_{n-1}(z)C_1^*\right|^2}{|h_n|^2}\notag,
\end{align}
where
\[
C_1 = \frac{3\Gamma+2i\Omega}{(\Gamma+i\Omega)^2}.
\]
The second term is given by,
\begin{align}
D_2^\pm &= \sum\limits_n \frac{\left(J_{n+1}(z)C_1\pm
J_{n-1}(z)C_1^*\right)}{h_n^*}\notag\\
&\times\left(\frac{J_{n+1}(z)C_2^*}{h_{n+1}^*}\pm\frac{
J_{n-1}(z)C_2}{h_{n-1}^*}\right)\notag,
\end{align}
with
\[
C_2 = \frac{2\Gamma+i\Omega}{(\Gamma+i\Omega)^2}.
\]
The final term is given by,
\begin{align}
D_3^\pm &= \sum\limits_n C_3
J_{n}^2(z)\left(\frac{1}{h_{n}^*}+\frac{1}{h_{n}}\right)\nonumber \\
& \pm C_4
J_{n-1}(z)J_{n+1}(z)\left(\frac{1}{h_{n-1}^*}+\frac{1}{h_{n+1}}\right)\notag
\end{align}
where
\begin{align}
C_3 &= \frac{5\Gamma^2+2\Omega^2}{(\Gamma^2+\Omega^2)^2}\notag\\
C_4 &= \frac{\Gamma C_2^2}{\Gamma+i2\Omega}+ \frac{\Gamma^2
}{(\Gamma+i\Omega)^2(\Gamma+i2\Omega)^2}.\notag
\end{align}
The effective damping and
diffusion can now be used to calculate the linewidth. These
quantities can in addition be integrated to give an effective
potential, allowing a calculation of the resonator
distribution.\cite{rodrigues:09} We note that this result
generalizes the phase diffusion rate calculated in Ref.\
\onlinecite{bennett:06} to the regime where $\Omega/\Gamma$ is no longer
small.

\end{document}